\documentclass[fleqn,10pt]{wlscirep}
\usepackage[utf8]{inputenc}
\usepackage[T1]{fontenc}
\usepackage{hyperref}

\usepackage{color} %% colored text
\definecolor{purple}{rgb}{0.63,0,1}
\definecolor{pink}{rgb}{1,0,0.9}

\DeclareMathOperator*{\argmin}{arg\,min}

\DeclareMathOperator{\tr}{tr}

\title{Application of Langevin Dynamics to Advance the Quantum Natural Gradient Optimization Algorithm}

\author[1,*]{Oleksandr Borysenko}
\author[1]{Mykhailo Bratchenko}
\author[1]{Ilya Lukin}
\author[2]{Mykola Luhanko}
\author[2]{Ihor Omelchenko}
\author[1,2]{Andrii Sotnikov}
\author[3]{Alessandro Lomi}

\affil[1]{National Science Center "Kharkiv Institute of Physics and Technology", Akademichna str. 1, 61108 Kharkiv, Ukraine}
\affil[2]{V.N. Karazin Kharkiv National University, Svobody Sq. 4, 61022 Kharkiv, Ukraine}
\affil[3]{Università della Svizzera italiana, Via Buffi 13, 6900 Lugano, Switzerland}

\affil[*]{alessandro.borisenko@gmail.com}

\begin{abstract}
A Quantum Natural Gradient (QNG) algorithm for optimization of variational quantum circuits has been proposed recently. In this study, we employ the Langevin equation with a QNG stochastic force to demonstrate that its discrete-time solution gives a generalized form of the above-specified algorithm, which we call Momentum-QNG. Similar to other optimization algorithms with the momentum term, such as the Stochastic Gradient Descent with momentum, RMSProp with momentum and Adam, Momentum-QNG is more effective to escape local minima and plateaus in the variational parameter space and, therefore, demonstrates an improved performance compared to the basic QNG. In this paper we benchmark Momentum-QNG together with the basic QNG, Adam and Momentum optimizers and explore its convergence behaviour. Among the benchmarking problems studied, the best result is obtained for the quantum Sherrington-Kirkpatrick model in the strong spin glass regime. Our open-source code is available at \url{https://github.com/borbysh/Momentum-QNG}
\end{abstract}
\begin{document}

\flushbottom
\maketitle
\thispagestyle{empty}

\section{Introduction}\label{intro}

Optimization of variational quantum circuits in hybrid quantum-classical algorithms has become a popular task over the recent time. The best known applications include the Variational Quantum Eigensolver (VQE) \cite{peruzzo2014variational}, Quantum Approximate Optimization Algorithm (QAOA) \cite{farhi2014quantum} and Quantum Neural Networks (QNNs) \cite{farhi2018classification, Huggins_2019, Schuld_2020}.

A computationally efficient method for evaluating analytic gradients on quantum hardware has been recently proposed \cite{PhysRevA.99.032331}. Therefore, the application of optimization algorithms from the Stochastic Gradient Descent (SGD) family has become possible. However, the path of steepest descent in the parameter space, guided by the (opposite) gradient vector direction, is usually not optimal, because it depends on the number of variational parameters, which is usually excessive. The same overparametrization problem is present in classical Machine Learning (ML) \cite{NIPS2015_eaa32c96}. To mitigate it, a Natural Gradient (NG) concept has been proposed \cite{10.1162/089976698300017746}. Contrary to vanilla SGD, NG determines the steepest descent direction taking into account the Fisher Information Matrix, which consists of the components 
 of the Riemannian metric tensor in the space of variational parameters. In this way, the optimization path becomes invariant under arbitrary reparametrization \cite{10.1162/089976698300017746} and, therefore, does not suffer from overparametrization \cite{pmlr-v89-liang19a}. To speed up calculations, the Fisher Information Matrix is usually approximated with different methods (see e.g. Refs.~\cite{pmlr-v37-martens15, Goldshlager_2024}).

%\section{Quantum Natural Gradient optimization algorithm}
Inspired by the NG approach, Stokes et al. \cite{Stokes_2020} have recently proposed its generalization to the quantum circuit optimization task, which they called the Quantum Natural Gradient (QNG) optimizer. They considered a parametric family of unitary operators $U_\theta \in U(D)$, which are indexed by real parameters $\theta \in \mathbb{R}^d$. 
With a fixed reference unit vector $|0\rangle \in \mathbb{C}^D$ and a Hermitian operator $H = H^\dag$ acting on $\mathbb{C}^D$, they consider the following optimization problem \cite{Stokes_2020}:
\begin{equation}\label{e:optim}
	\min_{\theta \in \mathbb{R}^d} \mathcal{L}(\theta) \, , 
    \quad 
    \mathcal{L}(\theta) = \frac{1}{2}\tr(P_{\psi_\theta} H) = \frac{1}{2}\langle \psi_\theta , H \psi_\theta \rangle \, ,
\end{equation}
where $\psi_\theta = U_\theta |0\rangle$ and $P_{\psi_\theta} \in \mathbb{CP}^{D-1}$ is the associated projector. Note that $\psi_\theta$ is normalized, since $U_\theta$ is unitary.  Any local optimum of the nonconvex objective function $\mathcal{L}(\theta)$ can be found by iterating the following discrete-time dynamical system \cite{Stokes_2020},
\begin{eqnarray}\label{e:normball}
	\theta_{n+1} = \argmin_{\theta \in \mathbb{R}^d} \left[\langle \theta - \theta_n, \nabla \mathcal{L}(\theta_n) \rangle
     + \frac{1}{2\eta} \Vert \theta - \theta_n \Vert_{g(\theta_n)}^2 \right] \, ,
\end{eqnarray}
where $\eta> 0$ is a positively defined constant, $g(\theta_n)$ is the symmetric matrix of the Fubini-Study metric tensor $g_{ij}(\theta) = \operatorname{Re}[G_{ij}(\theta)]$, and the Quantum Geometric Tensor is defined as follows (for further details see Stokes et al.\cite{Stokes_2020} and references therein):
\begin{equation}\label{e:qgt}
	G_{ij}(\theta) = \left\langle \frac{\partial \psi_\theta}{\partial \theta^i}  , \frac{\partial \psi_\theta}{\partial \theta^j} \right\rangle - \left\langle \frac{\partial \psi_\theta}{\partial \theta^i} , \psi_\theta\right\rangle  \left\langle \psi_\theta , \frac{\partial \psi_\theta}{\partial \theta^j}  \right\rangle  \, + \lambda \delta_{ij},
\end{equation}
where the last term with the Kronecker delta $\delta_{ij}$ and $\lambda \geq 0$ is added for regularization purpose.

Note, that generally the Quantum Geometric tensor can be defined for more general parametrized normalized wave functions. In particular, it naturally appears in the stochastic reconfiguration method \cite{sorella1998green} to optimize general variational quantum states and was used in the optimization of Neural Quantum States \cite{Carleo_2017, Chen_2024} and with certain modifications of some Tensor Network states \cite{puente2025efficient}. Besides, such tensors appear in the imaginary time evolution of Gaussian and generalized Gaussian states \cite{hackl2020geometry}. 

In equation~\eqref{e:normball}, according to Stokes et al.\cite{Stokes_2020}, the following notation is introduced:
\begin{equation}
	\Vert \theta - \theta_n \Vert_{g(\theta_n)}^2 = \langle \theta - \theta_n, g(\theta_n)(\theta - \theta_n) \rangle \enspace .
\end{equation}
Then, the first-order optimality condition corresponding to equation~\eqref{e:normball} is:
\begin{equation}\label{eq:linearsystem}
	g(\theta_n)(\theta_{n+1} - \theta_n) = -\eta \cdot \nabla \mathcal{L}(\theta_n) \enspace .
\end{equation}
A solution of the optimization problem~\eqref{e:normball} is thus provided by the following expression which involves the inverse $g^{-1}(\theta_n)$  of the metric tensor:
\begin{equation}\label{eq:qgtoptimization}
	\theta_{n+1} - \theta_n = - \eta \cdot  g^{-1}({\theta_n}) \nabla \mathcal{L}(\theta_n) \enspace .
\end{equation}

In this way, with the Fubini-Study metric tensor introduced, the implied descent direction in the parameter space, given by the right-hand side of equation~\eqref{eq:qgtoptimization}, becomes invariant with respect to arbitrary reparametrization and, therefore, to the details of the quantum circuit architecture under consideration. 
With their QNG optimizer~\eqref{eq:qgtoptimization}, the authors \cite{Stokes_2020} achieved a considerable improvement in optimization performance compared to SGD and Adam \cite{kingma2017adam}. 

However, being rather effective for convex optimization, QNG often sticks to local minima, saddles and plateaus of nonconvex loss functions. In classical ML applications, employing optimization algorithms with momentum (inertial) term, such as SGD with momentum \cite{backpropagation}, RMSProp with momentum \cite{tieleman2012lecture} and Adam \cite{kingma2017adam}, has demonstrated better convergence characteristics. 

Recently, Borysenko and Byshkin \cite{coolmomentum} demonstrated that a discrete-time solution of the Langevin equation with stochastic gradient force term results in the well-known  SGD with momentum \cite{backpropagation} optimization algorithm. In this paper, we study a particular case of Langevin dynamics with the QNG stochastic force term. In Section \ref{Langevin} we give a brief picture of the multidimensional discrete-time Langevin dynamics and show its relation to the stochastic optimization process. Based on these results, in Section \ref{Results} we derive a generalized QNG optimization algorithm, which we call Momentum-QNG and benchmark it together with the basic QNG, Momentum and Adam on several optimization tasks to demonstrate its improved performance.

\section{Langevin dynamics and its relation to stochastic optimization} \label{Langevin}
The adaptation of Langevin dynamics for optimization suggests a new prospective research direction \cite{NIPS2017_6664, ma2019sampling}. For optimization of analytically defined objective functions, even a quantum form of the Langevin dynamics has been proposed \cite{Quantum_Langevin}.

In Langevin dynamics, two forces are added to the classical Newton equation of motion: a viscous friction force proportional to the velocity with a friction coefficient $\gamma \geq 0$ and a thermal white noise. Explicitly, the Langevin dynamics of a (virtual) Brownian particle with unit mass $m=1$ in the space of real variables $\theta \in \mathbb{R}^d$ and real time $t$ can be described by the following  equation (see e.g. Refs.\cite{bussi2007accurate,vanden2006second, van1982algorithms,schlick2010molecular}):

\begin{equation} \label{eq:Langevin}
\frac{dv(t)}{dt}= f(\theta)-\gamma v(t)+R(t), 
\end{equation}
where $v=d\theta/dt$ denotes velocity, $f(\theta)\in \mathbb{R}^d$ is a regular force and $R(t)\in \mathbb{R}^d$ is a random uncorrelated force with zero mean $\left \langle R(t) \right \rangle=0$ and temperature-dependent magnitude:

\begin{equation} \label{eq:Thermostat}
\left \langle R^{i}(t) \cdot R^{j}(t') \right \rangle=2T\gamma \delta (t-t') \delta_{ij}, 
\end{equation} 
$\delta (t-t')$ being the Dirac delta function.

Similar to the $d=3$ case, the temperature $T$ in the $d$-dimensional space of our variational parameters may be introduced as twice the mean kinetic energy $E_{k}$ of a particle divided by $d$:
\begin{equation} \label{eq:Temperature}
T = \frac{2E_{k}}{d} = \frac{1}{d}\sum_{i=1}^{d} \left \langle \left(v^i\right)^2 \right \rangle.
\end{equation}
In equation~\eqref{eq:Temperature} we set the Boltzmann's constant $k_{B}=1$ for brevity.

The discrete-time form of equation~\eqref{eq:Langevin} with stochastic force $\hat{f}=f+R$ reads: 
\begin{equation} \label{eq:Langevin_fin}
    \frac{\Delta \theta_{n+1} - \Delta \theta_{n}}{\Delta t^{2}} = \hat{f}_{n} - \gamma \frac{\Delta \theta_{n+1} + \Delta \theta_{n}}{2 \Delta t},
\end{equation}
where $\Delta \theta_{n+1} = \theta_{n+1} -  \theta_{n}$ and $\Delta t$ is a time step.
Now, it is straightforward to obtain the next parameter updating formula:
\begin{equation} \label{eq:momentum}
    \Delta \theta_{n+1} = \rho \Delta \theta_{n} + \hat{f}_{n} \cdot \eta 
\end{equation}
with
\begin{equation} \label{eq:rho}
    \rho = \frac{1-\gamma \Delta t /2}{1+\gamma \Delta t /2}
\end{equation}
and
\begin{equation} \label{eq:lr}
    \eta = \frac{\Delta t^{2}}{1+\gamma \Delta t /2} = \frac{1+\rho}{2}\Delta t^{2}.
\end{equation}

Equation~\eqref{eq:momentum} is nothing else but a well-known SGD with momentum optimization algorithm \cite{backpropagation} (further referred to as Momentum) with $\rho$ being a momentum coefficient and $\eta$ a learning rate constant.

After changing to discrete time, equation~\eqref{eq:Thermostat} becomes:
\begin{equation} \label{eq:Thermostat_gamma}
\left \langle R^{2} \right \rangle \Delta t = 2T\gamma, 
\end{equation} 
where $\left \langle R^{2} \right \rangle = d^{-1} \cdot \sum_{i=1}^{d}\left \langle \left(R^{i}\right)^{2} \right \rangle$.

Using equation~\eqref{eq:rho} to change variables from $\gamma$ to $\rho$, the thermostatic condition \eqref{eq:Thermostat_gamma}, corresponding to the discrete-time form of the Langevin equation \eqref{eq:Langevin_fin}, becomes:
\begin{equation} \label{eq:Thermostat_rho_t}
\left \langle R^{2} \right \rangle \Delta t^{2} = 4T\cdot \frac{1-\rho}{1+\rho}.
\end{equation}

At the same time, in discrete variables, the temperature \eqref{eq:Temperature} becomes
\begin{equation} \label{eq:T}
T = \frac{1}{d}\sum_{i=1}^{d} \left \langle\ \left( \frac{ \Delta\theta^i }{\Delta t} \right)^2 \right \rangle = \frac{1}{d \cdot \Delta t^2}\sum_{i=1}^{d} \left(\sigma^i\right)^2 = \frac{\sigma^2}{ \Delta t^2},
\end{equation}
where $\sigma=\sqrt{d^{-1} \cdot \sum_{i=1}^{d} \left(\sigma^i\right)^2}$ is a standard deviation magnitude in the space of variational parameters.

Now, with  equations~\eqref{eq:lr} and \eqref{eq:T} in mind, one can derive from the thermostatic condition~\eqref{eq:Thermostat_rho_t} the next relation:
\begin{equation} \label{eq:sigma}
\sigma= \eta \cdot \sqrt\frac{ \left \langle R^{2} \right \rangle}{1-\rho^2},
\end{equation}
which describes fluctuations of variational parameters in the vicinity of a loss function extremum under the influence of random noise. One can use equation~\eqref{eq:sigma} to estimate the accuracy of the optimized variational parameters.

From equation~\eqref{eq:sigma} one can see that the mean random jump length $\sigma$ increases as the momentum coefficient $\rho$ increases. Therefore, choosing $\rho>0$ helps to escape local minima and plateaus, where the local gradient values vanish.

\section{Benchmarking of the quantum natural gradient descent with momentum} \label{Results}

Setting
\begin{equation} \label{eq:qng_force}
    \hat{f}_n=- g^{-1}({\theta_n}) \nabla \mathcal{L}(\theta_n),
\end{equation}
from equation~\eqref{eq:momentum} one obtains a generalized form of the QNG optimizer~\eqref{eq:qgtoptimization}:
\begin{equation} \label{eq:momentum-qng}
    \Delta \theta_{n+1} = \rho \Delta \theta_{n} - \eta \cdot g^{-1}({\theta_n}) \nabla \mathcal{L}(\theta_n),      
\end{equation}
which we call Momentum-QNG. Note, that equation~\eqref{eq:momentum-qng} reduces to the basic QNG~\eqref{eq:qgtoptimization} for $\rho = 0$.

From equation~\eqref{eq:momentum-qng} one can see that our Momentum-QNG optimization algorithm is a quantum adaptation of SGD with momentum \cite{backpropagation} and should not be confused with the recently introduced Momentum QNG algorithm \cite{Fitzek2024optimizing} (which is a quantum adaptation of Adam \cite{kingma2017adam}) proposed by the authors of qBang \cite{Fitzek2024optimizing}.

In this section we benchmark four gradient-based optimization algorithms, integrated into the PennyLane \cite{bergholm2022pennylaneautomaticdifferentiationhybrid} quantum computation package as Python3 classes, at different learning rate values with the rest hyperparameter values set as follows: Adam($\beta_1=0.9, \beta_2=0.99, \epsilon=10^{-8}$), Momentum($\rho=0.9$), QNG($\lambda=0.5$), Momentum-QNG($\rho=0.9, \lambda=0.5$). In QNG and Momentum-QNG, the Fubini-Study metric tensor (see equation~\eqref{e:qgt}) is computed under the block-diagonal approximation and the regularization coefficient $\lambda = 0.5$ is applied. In all calculations we set $\Delta \theta_0 = 0$ as the initial condition. 

\subsection{Variational Quantum Eigensolvers} \label{vqe}
\subsubsection{Investment Portfolio Optimization}

Our first test-drive model is an Investment Portfolio Optimization task, which can be mapped to the $N$-particle ($N$ is a number of companies in portfolio) Ising spin-glass model. The ground state of the corresponding $N$-qubit Hamiltonian is parametrized with the VQE ansatz and further optimized to find the energy minimum. Here we explore the cases $N=6$, 11 and 12. To benchmark the performance of different optimizers, we run a series of 200 trials on a modified tutorial code by Chi-Chun Chen \cite{portfolio_optimization_tutorial} for a range of learning rate values: $0.01 \leq \eta \leq 3$. Each trial is initialized with random-guess values of variational parameters, being the same for all optimizers. 
Next, the optimization process runs for 200 steps or until energy convergence up to the 3-digit accuracy. As a result of each optimization run, we calculate $\Delta E = E_{\text{opt}} - E_{\text{ground}}$ -- the difference between the optimized and the exact ground state energy. 

\begin{figure}[t]
\centering
    \includegraphics[width= 0.5\linewidth]{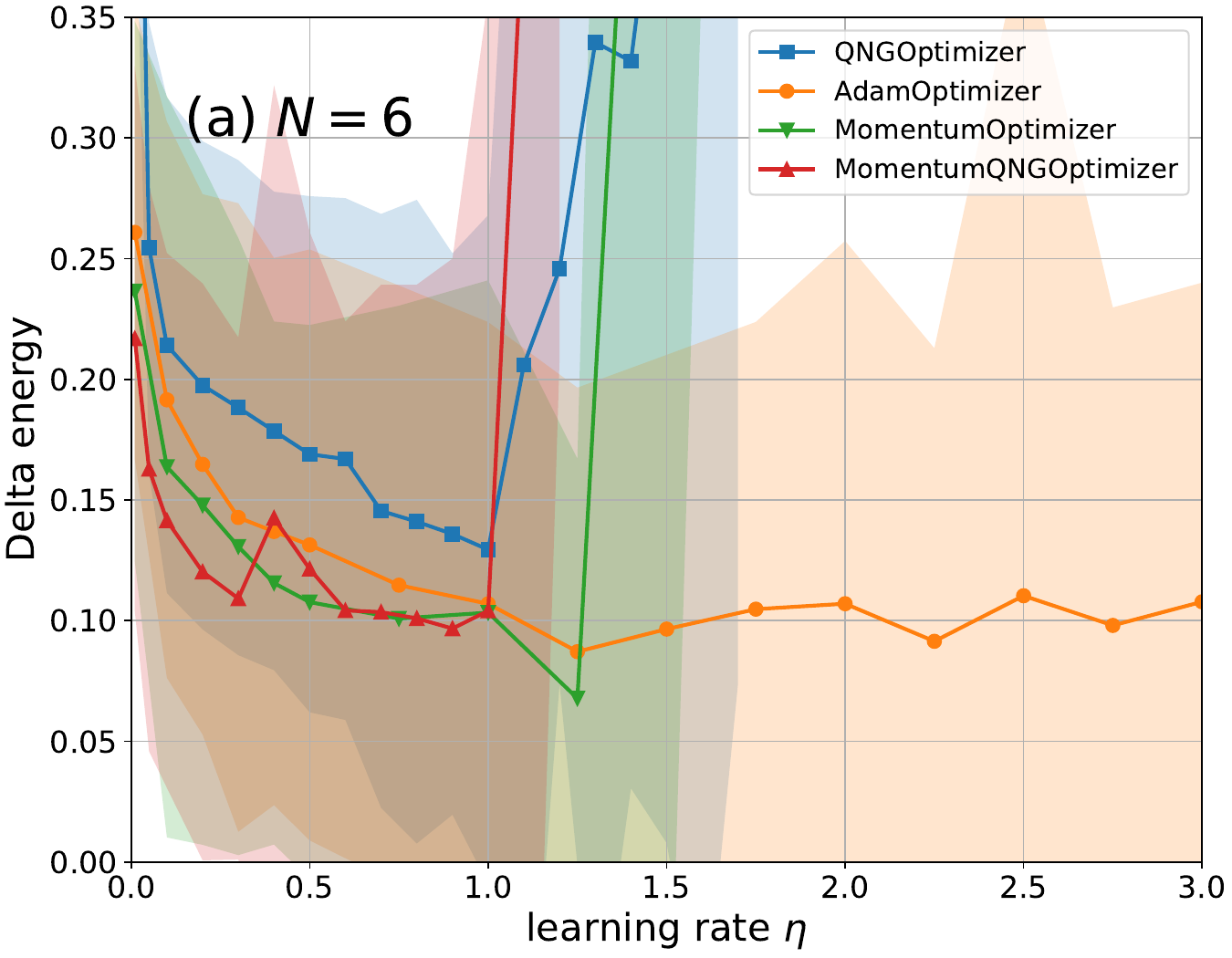}\includegraphics[width= 0.5\linewidth]{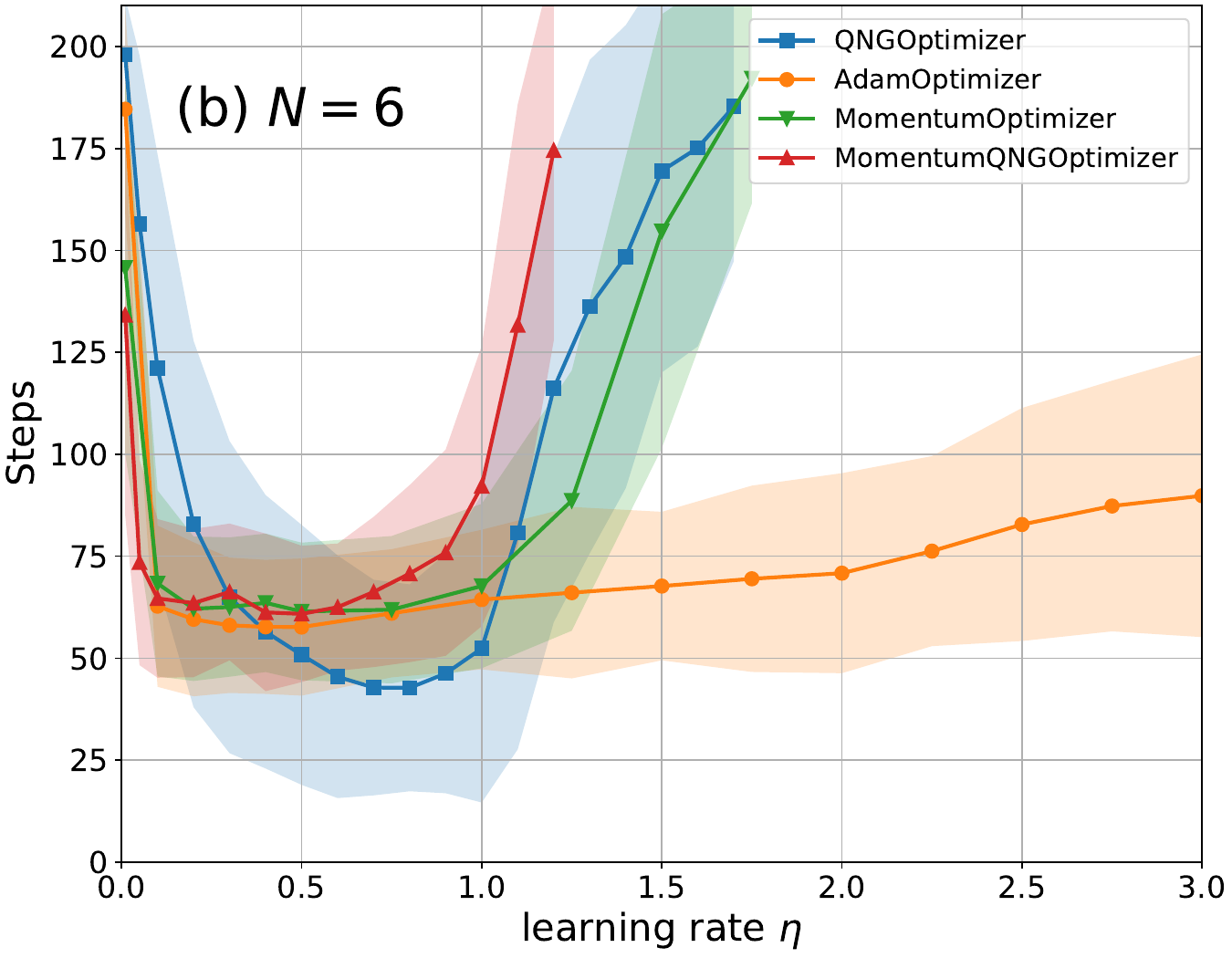}
    \includegraphics[width= 0.5\linewidth]{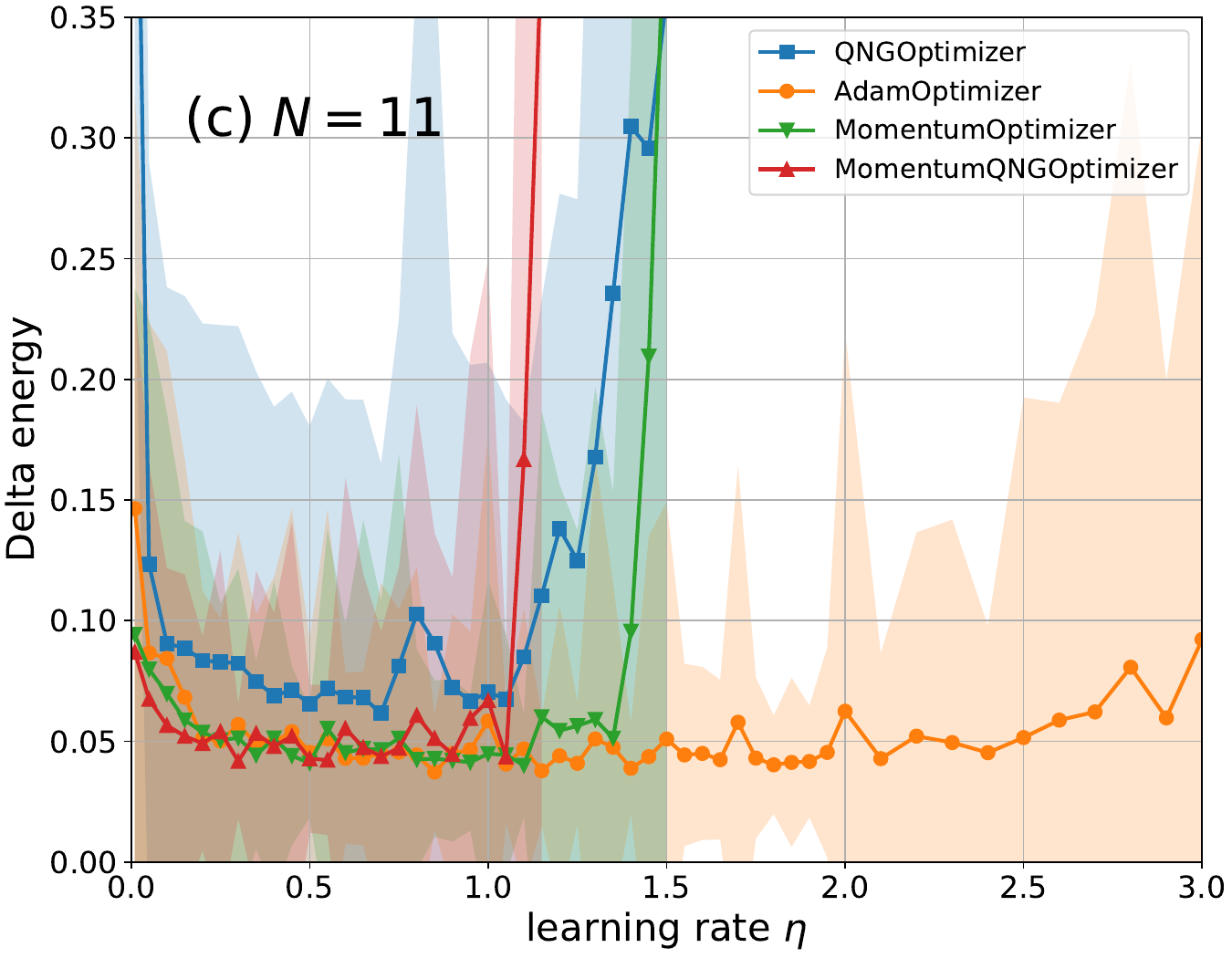}\includegraphics[width= 0.5\linewidth]{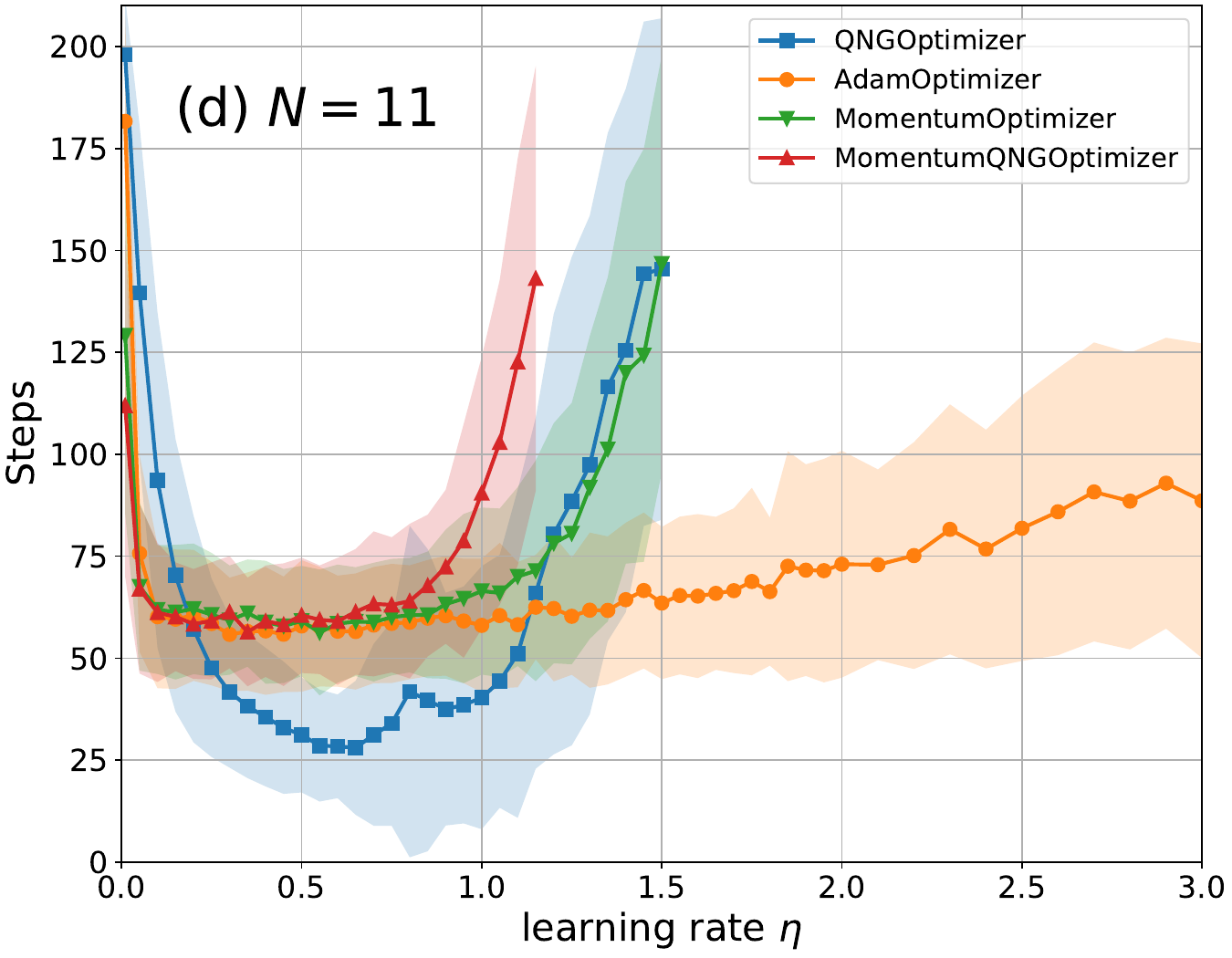}
    \includegraphics[width= 0.5\linewidth]{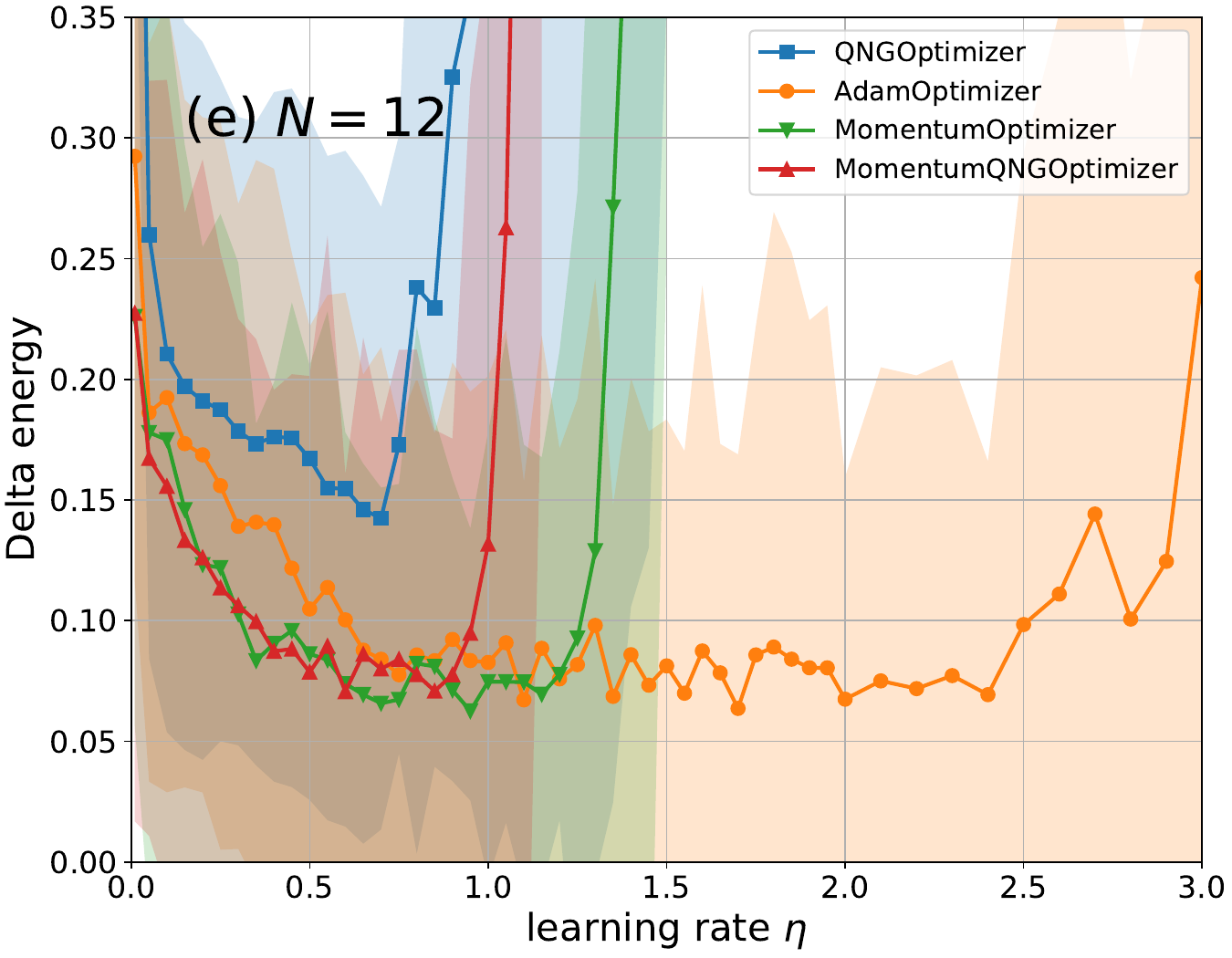}\includegraphics[width= 0.5\linewidth]{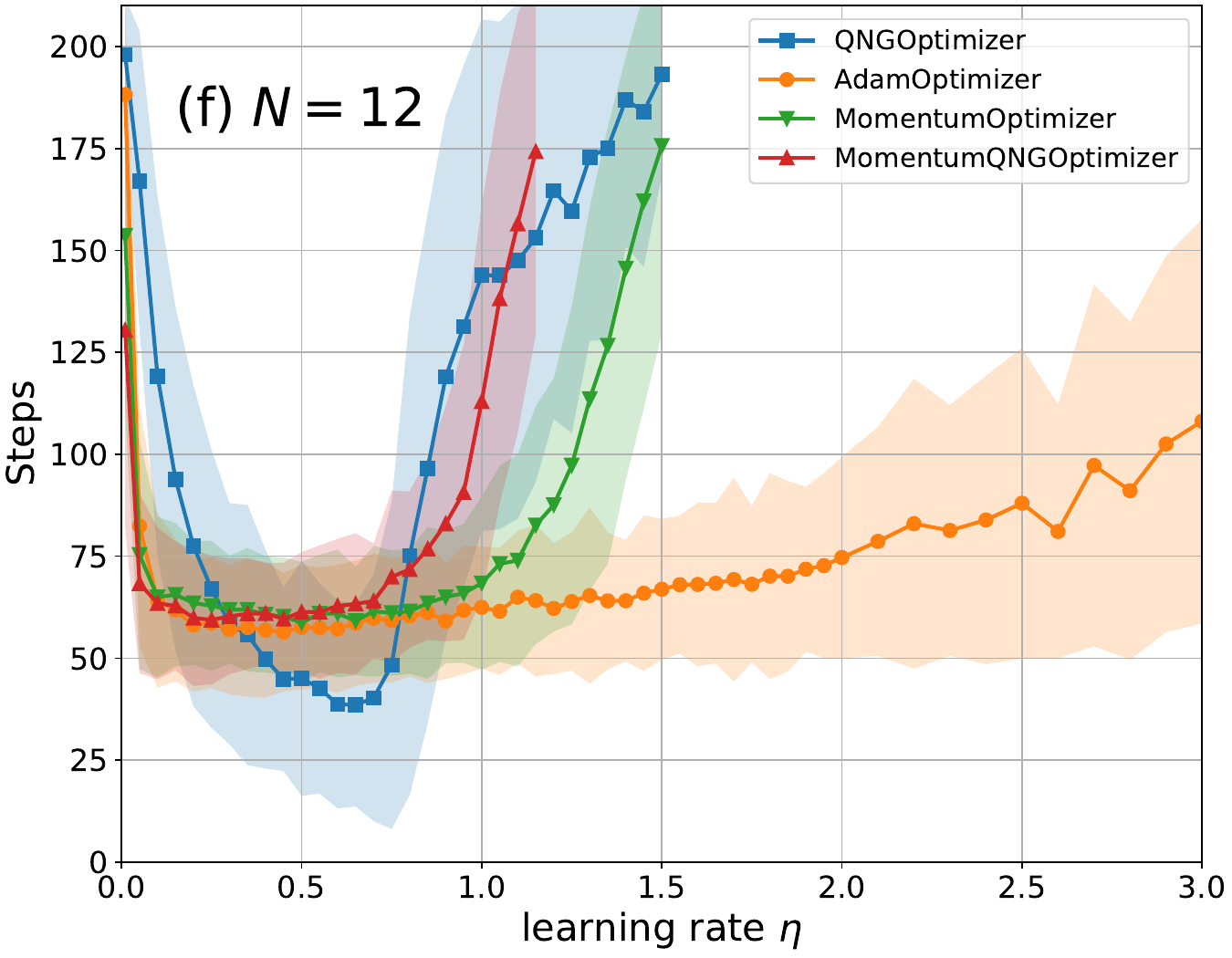}
    
    \caption{\label{fig:Portfolio}%
        Benchmarking Momentum-QNG together with QNG, Momentum and Adam on the portfolio optimization problem. The vertical axis shows the mean (symbols) and the standard deviation (shaded regions) of the difference between the optimized and the ground state energy (a) ($N=6$), (c) ($N=11$), (e) ($N=12$) and of the number of steps to convergence (b) ($N=6$), (d) ($N=11$), (f) ($N=12$) in a series of 200 trials, while the horizontal axis shows the learning rate $\eta$ of four different optimizers under consideration. 
        }
\end{figure}

To compare the performance of different optimizers, on Fig.~\ref{fig:Portfolio}(a), (c) and (e) we plot the mean (symbols) and standard deviation (shaded regions) values of $\Delta E$ as a function of the learning rate $\eta$.  
From Fig.~\ref{fig:Portfolio}(a), (c) and (e) one can see that for $N=6$, $N=11$ and $N=12$ all the three momentum-amended optimizers (Momentum-QNG, Momentum and Adam) give similar best results in their convergence domain and significantly outperform the momentumless QNG.

To study the convergence behaviour of the optimization algorithms under consideration, in Fig.~\ref{fig:Portfolio}(b), (d) and (f) we plot the mean (symbols) and the standard deviation (shaded regions) of the number of steps to convergence, as a function of the learning rate $\eta$. Again, for $N=6$, $N=11$ and $N=12$ all the momentum-amended optimizers demonstrate similar results in their convergence domain. The momentumless QNG shows the fastest convergence behaviour, though the highest energy misfit.

It is worth noting that Adam demonstrates the most robust performance with the widest convergence domain. 
For further details of our calculations see our code \cite{Borbysh_portfolio}.

\subsubsection{The Sherrington-Kirkpatrick model} \label{sk}

In this section we optimize the quantum Sherrington-Kirkpatrick (SK) model \cite{Sherrington-Kirkpatrick} to find its ground-state energy in the framework of the VQE approximation. For a recent discussion of the quantum SK model and its ground state ansatzes see Ref.\cite{Schindler2022}. 

We consider the  $N$-qubit quantum SK model in the transverse field, defined by the following Hamiltonian:

\begin{equation} \label{eq:SK_Ham}
    \hat{H} = \sum_{i,j} J_{ij}\hat{\sigma}_i^z \hat{\sigma}_j^z - g\sum_{i} \hat{\sigma}_i^x,      
\end{equation}
where the first sum is taken over all pairs of sites  $(ij)$  and  $J_{ij}=\mathcal{N}(0,1)/\sqrt{N}$  are sampled from the normal distribution with zero mean and  $1/N$  variance. At small  $g<1.5$  the model is generally in the spin glass phase. At large  $g$  the model becomes paramagnetic in the $x$ direction.

To illustrate the performance of different optimizers, we consider the VQE optimization problem aimed to minimize the ground state energy expectation value:

\begin{equation} \label{eq:E_0}
    E_0(\theta) =  \langle 0| U^{\dag}_{\theta} \hat{H} U_{\theta} |0  \rangle ,      
\end{equation}
where $U_{\theta}$ is a parametrized unitary matrix.

\begin{figure}[t]
\centering
    \includegraphics[width= 0.5\linewidth]{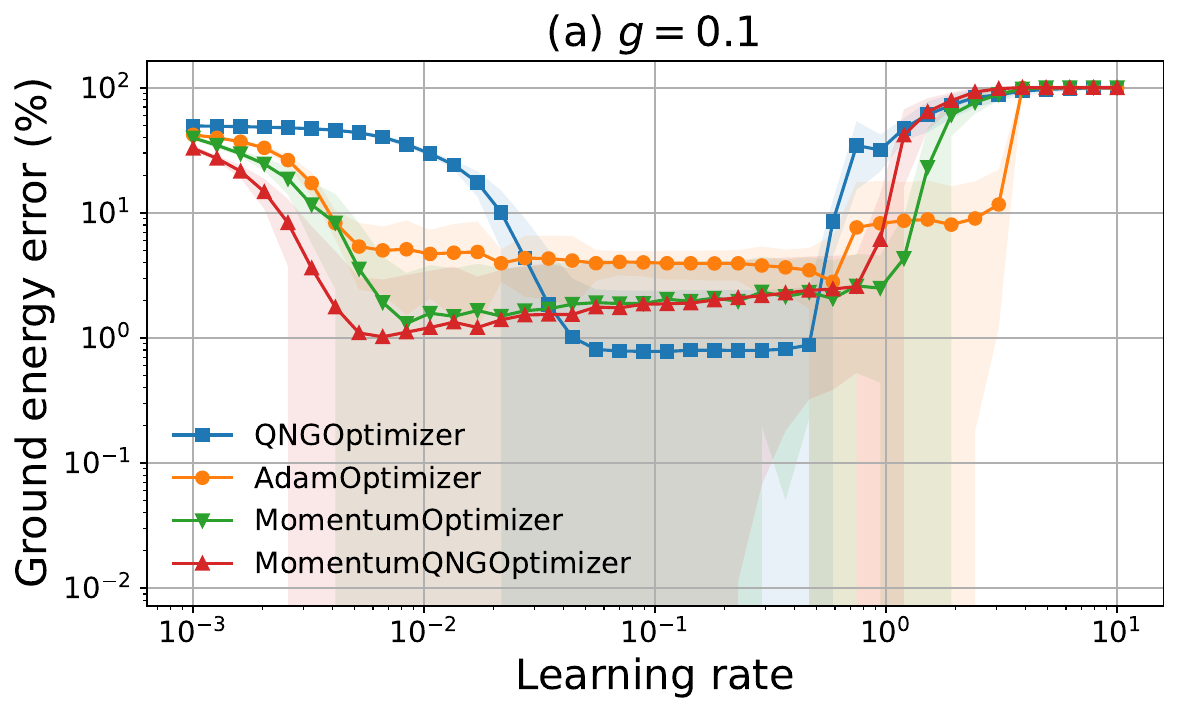}\includegraphics[width= 0.5\linewidth]{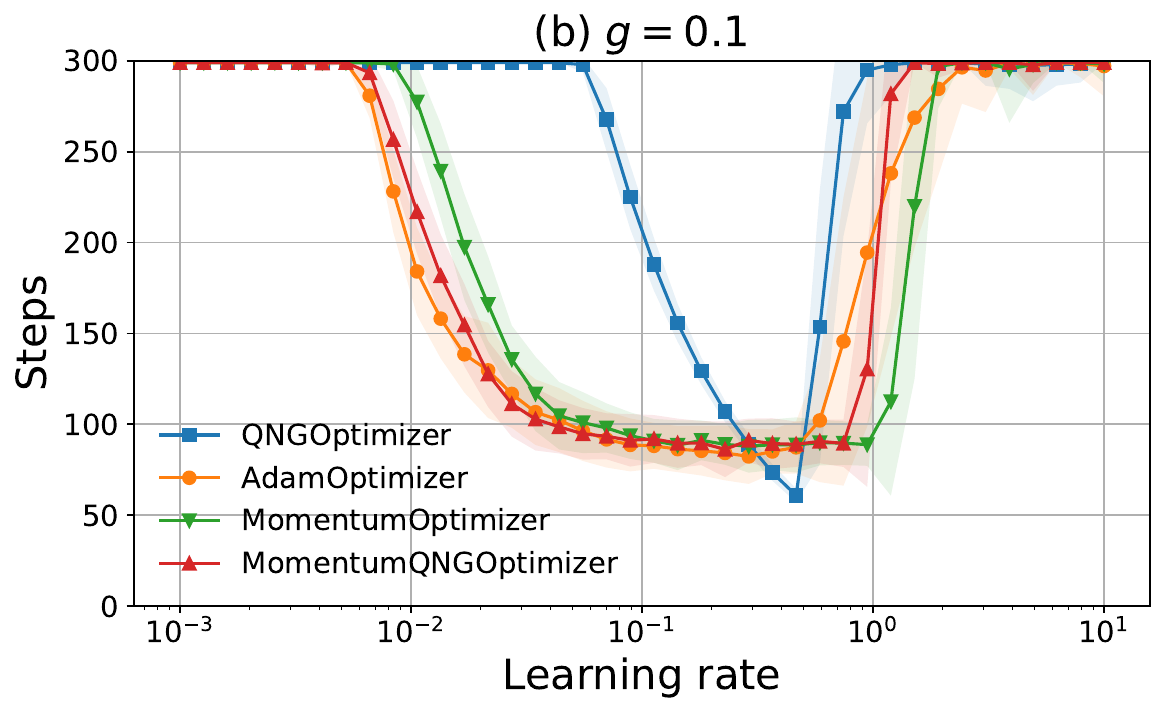}
    \includegraphics[width= 0.5\linewidth]{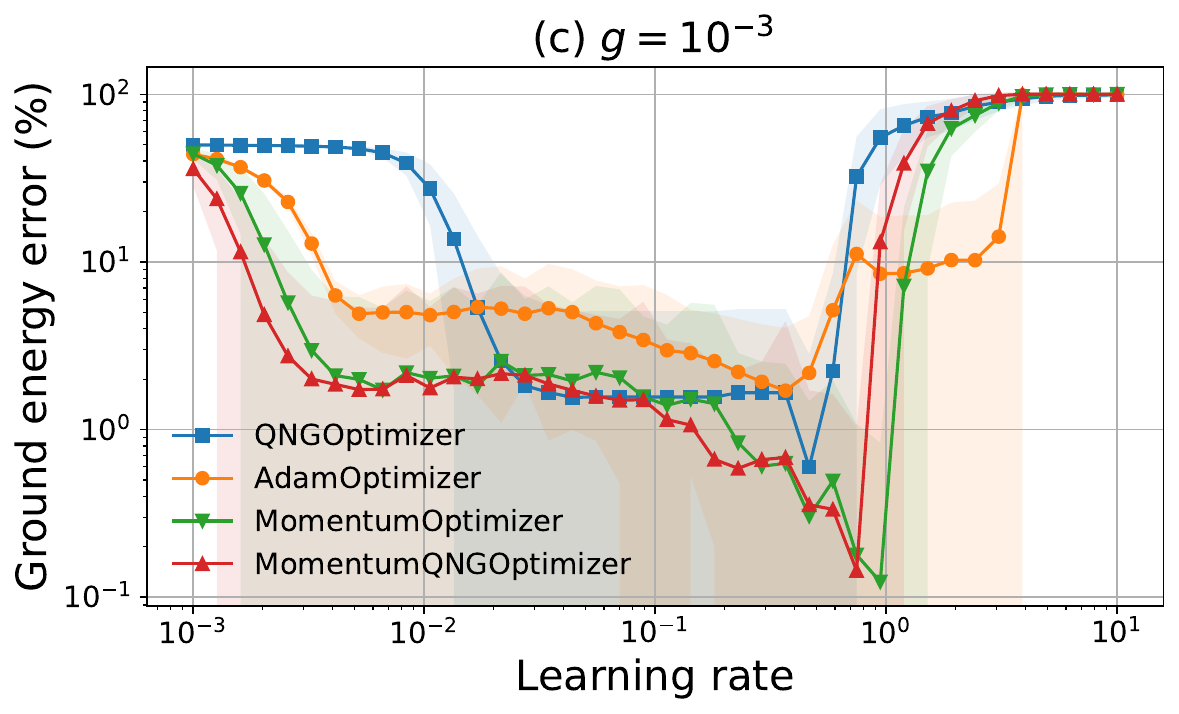}\includegraphics[width= 0.5\linewidth]{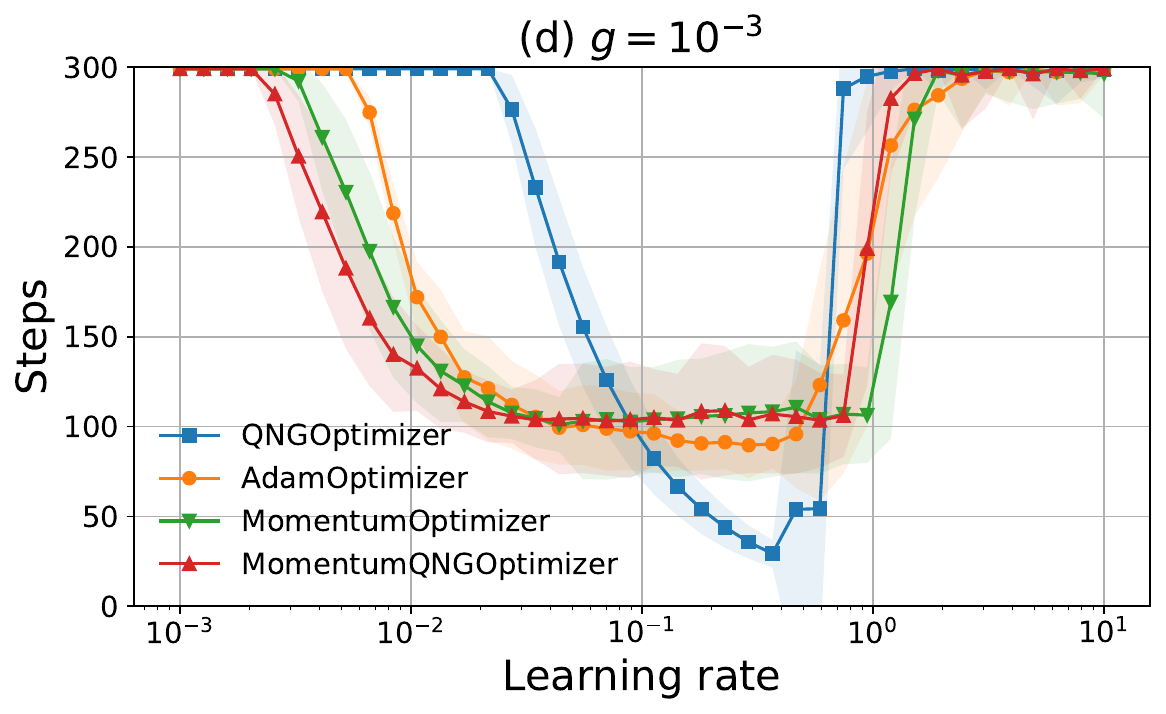}
    \includegraphics[width= 0.5\linewidth]{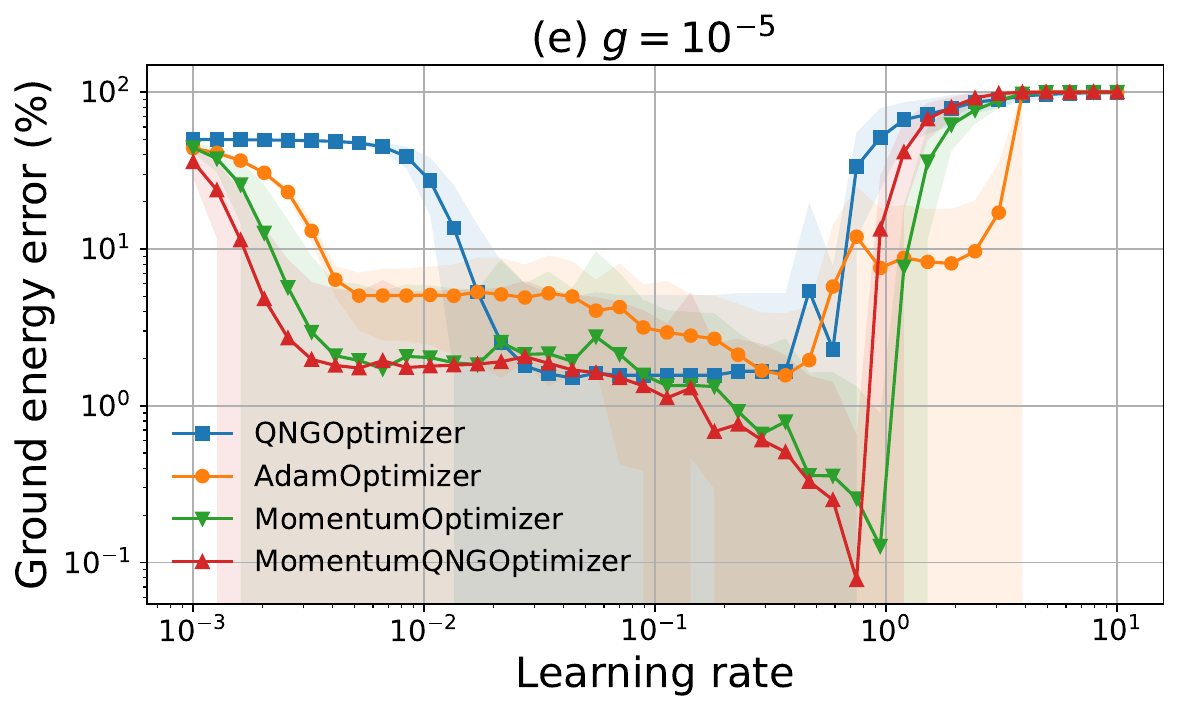}\includegraphics[width= 0.5\linewidth]{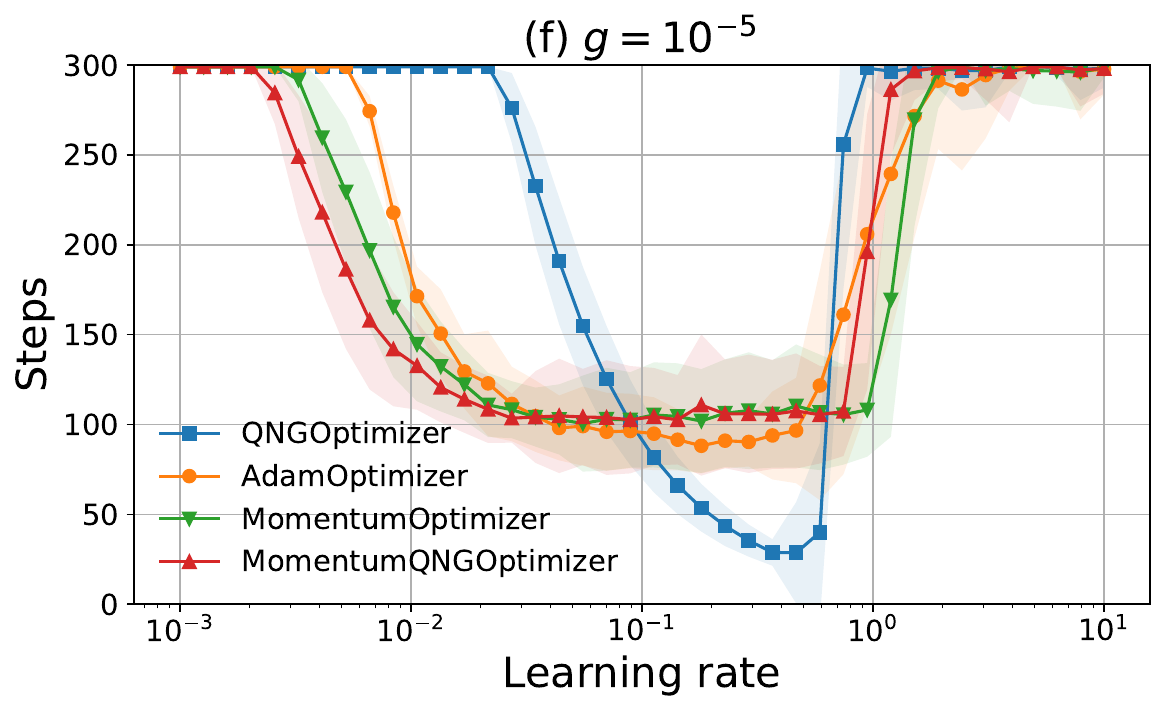}
    
    \caption{\label{fig:SK}%
        Benchmarking Momentum-QNG together with QNG, Momentum and Adam on the Sherrington-Kirkpatrick model at three different values of transverse field $g$ (indicated at figure captions). The vertical axis shows the mean (symbols) and the standard deviation (shaded regions) of the difference (in percents) between the optimized and the true ground state energy (a) ($g=0.1$), (c) ($g=10^{-3}$), (e) ($g=10^{-5}$) and of the number of steps to convergence (b) ($g=0.1$), (d) ($g=10^{-3}$), (f) ($g=10^{-5}$) in a series of 200 trials, while the horizontal axis shows the learning rate $\eta$ of four different optimizers under consideration. 
        }
\end{figure}

In Fig.~\ref{fig:SK} below we demonstrate results for $N=8$ qubits. To benchmark the performance of different optimizers, we run a series of 200 trials based on our modified tutorial code \cite{SK_tutorial} for a range of learning rate values: $10^{-3} \leq \eta \leq 10$. Each trial is initialized with random-guess values of variational parameters, being the same for all optimizers. 
Next, the optimization process runs for 300 steps or until energy convergence up to the 5-digit accuracy. As a result of each optimization run, we calculate the $\text{error} = 100\% \cdot \left(E_{0}^{\text{true}} - E_{0}(\theta) \right)/E_{0}^{\text{true}}$ -- the relative difference between the optimized and the true ground state energy. 

To compare the performance of different optimizers, on Fig.~\ref{fig:SK}(a), (c) and (e) we plot the mean (symbols) and standard deviation (shaded regions) values of the ground energy error as a function of the learning rate $\eta$. From Fig.~\ref{fig:SK} one can see that for $g=0.1$ (a) the basic QNG achieves the least error value. For $g=10^{-3}$ (c) Momentum and Momentum-QNG demonstrate almost equal best results and for $g=10^{-5}$ (e) Momentum-QNG shows the best result. It is worth to note that Adam demonstrates modest optimization performance in all three cases.

To study the convergence behaviour of the optimization algorithms under consideration, in Fig.~\ref{fig:SK}(b), (d) and (f) we plot the mean (symbols) and the standard deviation (shaded regions) of the number of steps to convergence, as a function of the learning rate $\eta$. One can see that the basic QNG demonstrates the fastest convergence in the most narrow domain, while the rest three optimizers behave similarly.

Our raw data and additional numeric results can be found at our project page \cite{Borbysh_sk}.

\subsection{Quantum Approximate Optimization Algorithm} \label{qaoa}
Our next test model is the Minimum Vertex Cover problem treated in the framework of the QAOA approach. Recently, this problem has been used to study the impact of noise on classical optimizers and to determine the optimal depth of the QAOA circuit \cite{Pellow-Jarman2024-kv}. In our calculations we use a modified code by Jack Ceroni \cite{Ceroni2024} to study two graphs with $N=4$ and $N=8$ vertices. We build QAOA circuits with 4 layers for $N=4$ qubits and with 6 layers for $N=8$ qubits. 
Then we run a series of 200 trials with the same for all optimizers random-guessed initial values of variational parameters. The optimization process runs for 200 steps or until energy convergence up to the 2-digit accuracy during at least 3 steps. To compare the performance of different optimizers, we calculate the quality ratio of the final optimized state -- the total probability to find the states of the exact solution in the given optimized solution. The range of learning rate values studied is $0.01 \leq \eta \leq 2$ for $N=4$ and $0.001 \leq \eta \leq 0.9$ for $N=8$.

\begin{figure}[t]
\centering
    \includegraphics[width= 0.5\linewidth]{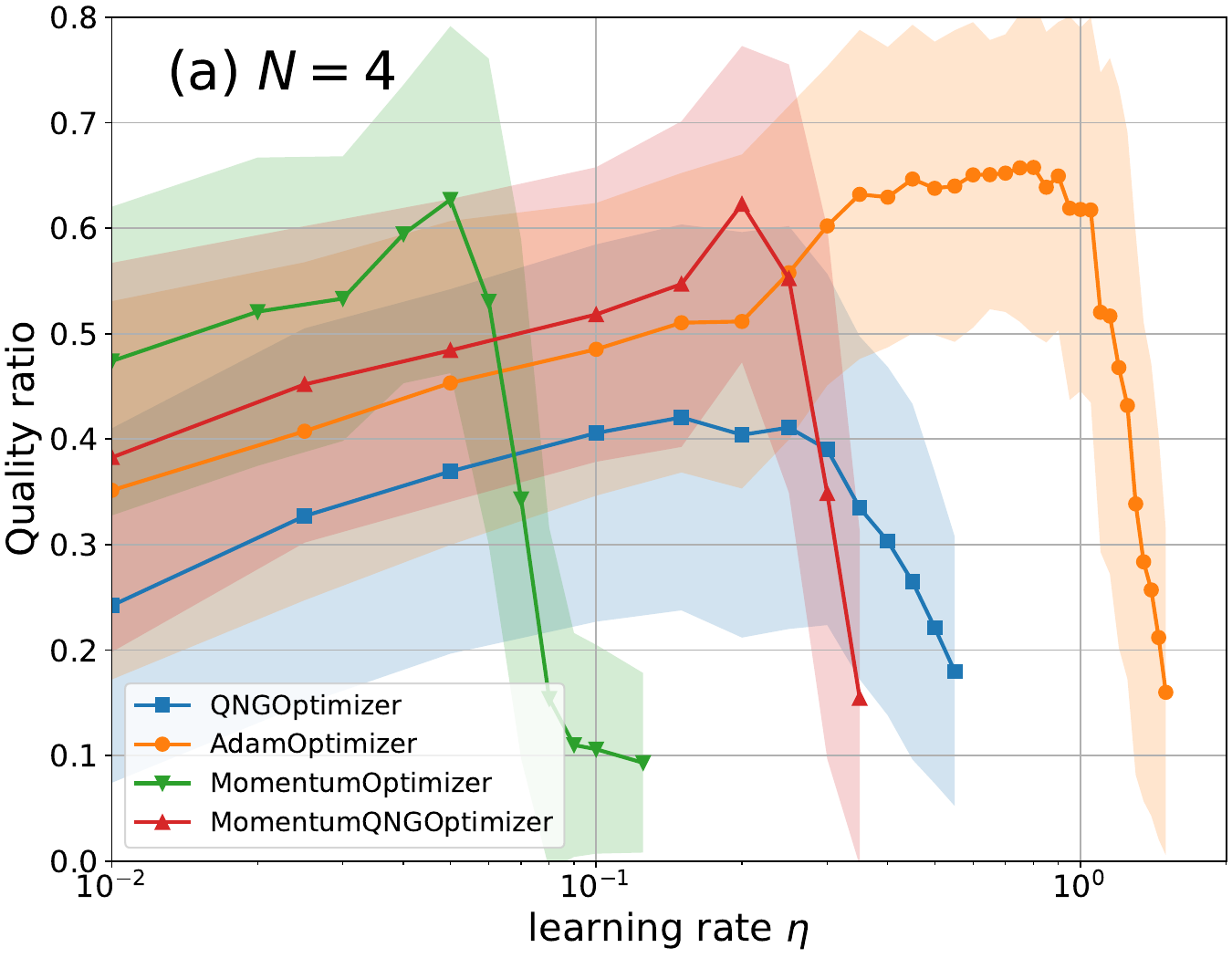}\includegraphics[width= 0.5\linewidth]{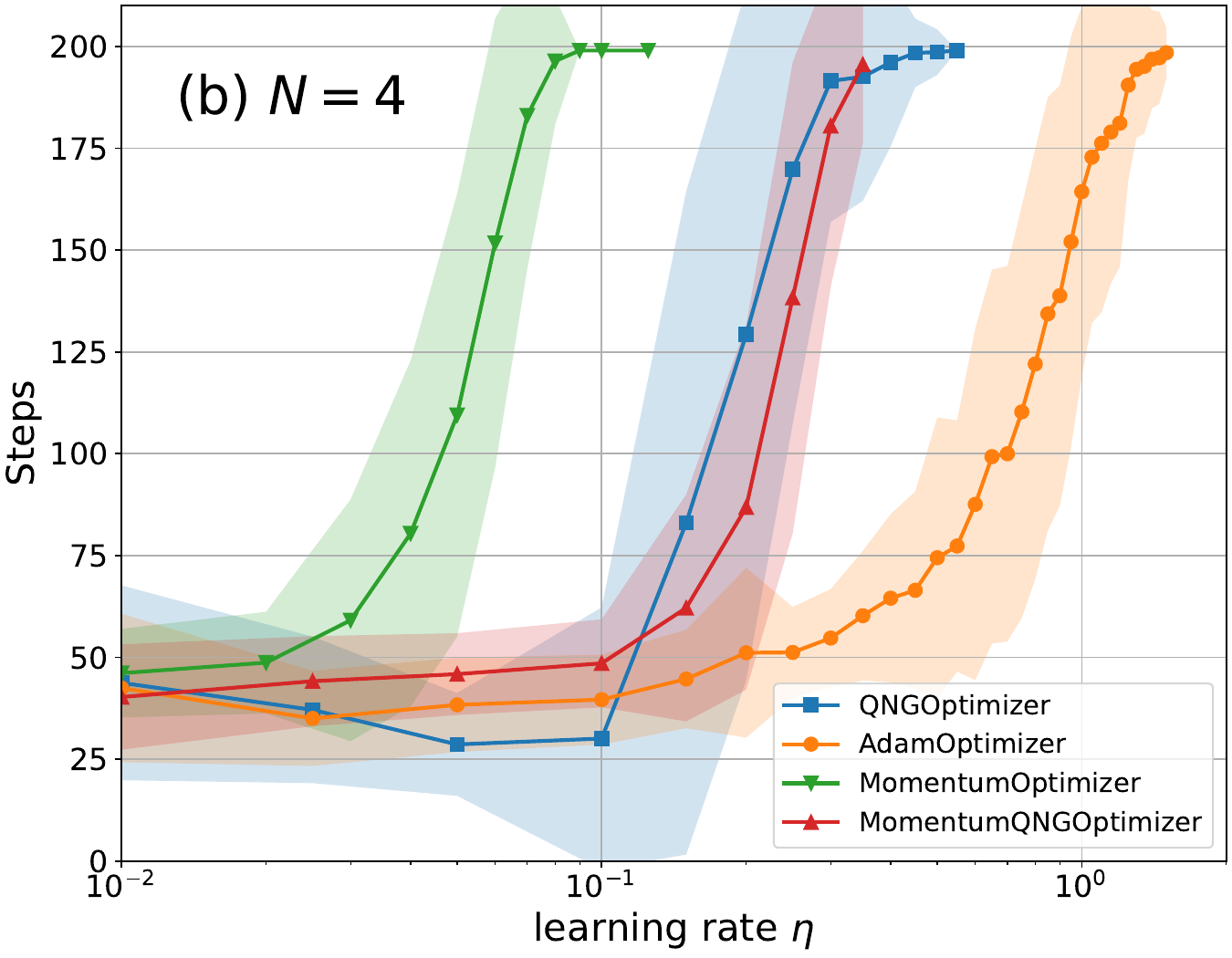}
    \includegraphics[width= 0.5\linewidth]{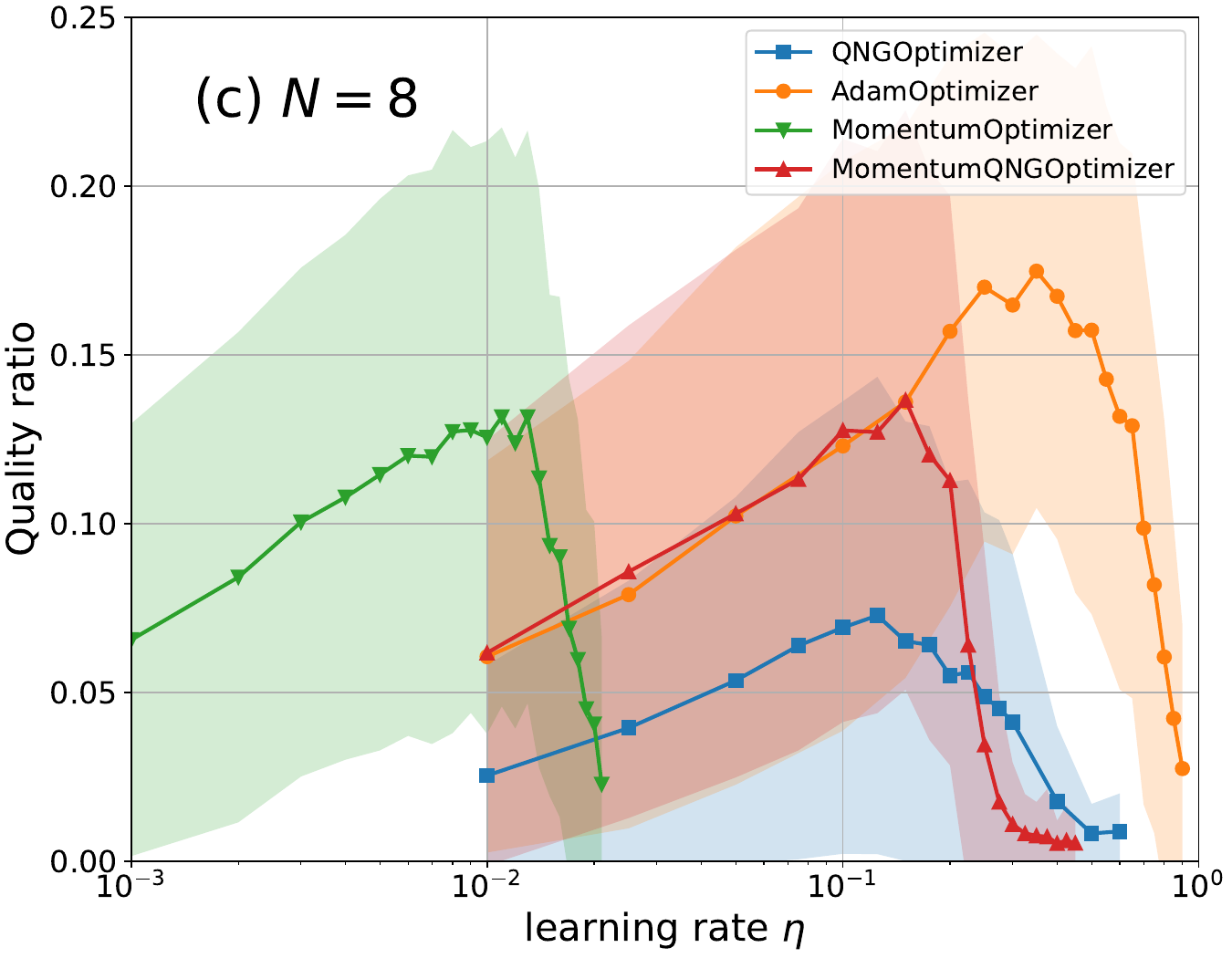}\includegraphics[width= 0.5\linewidth]{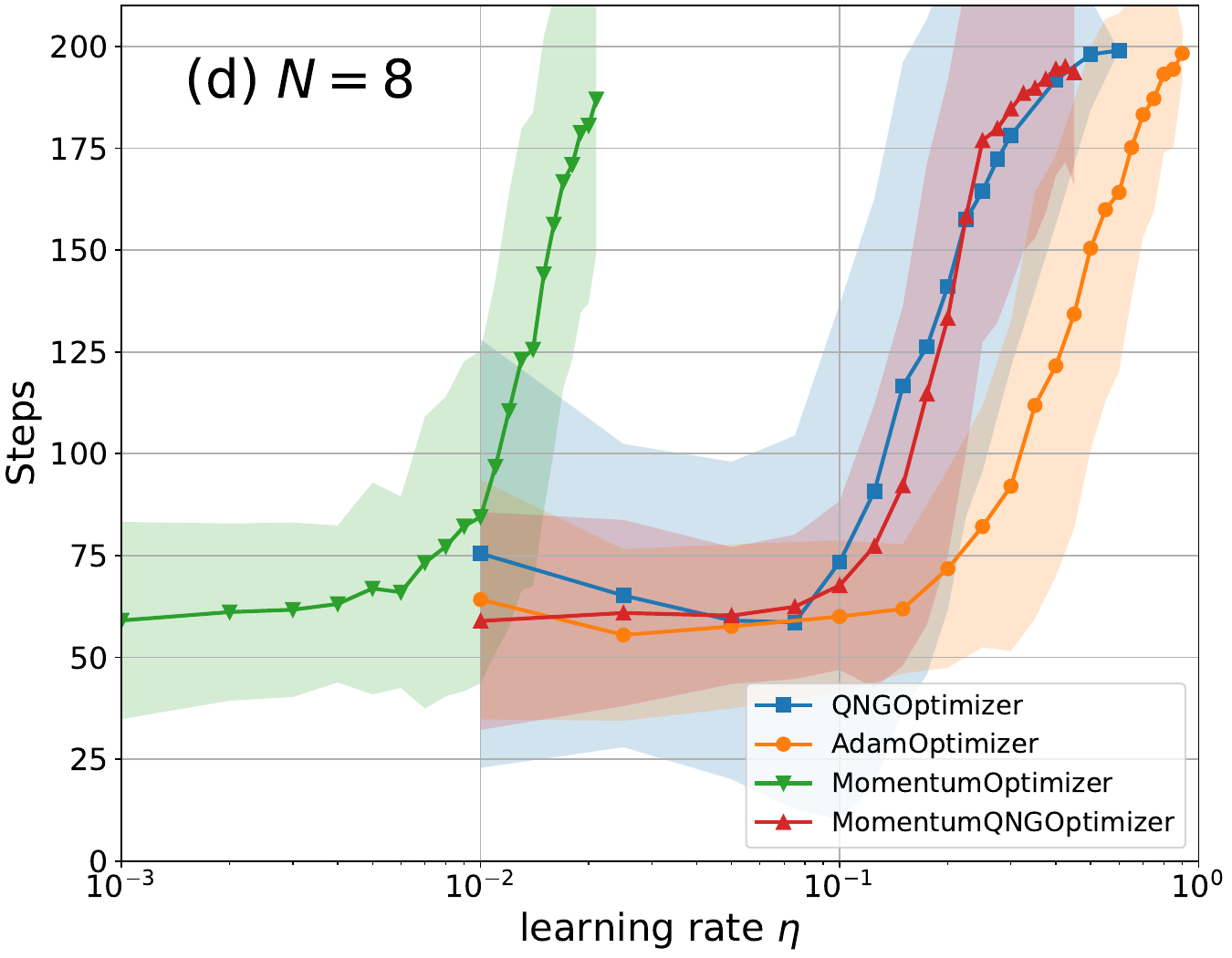}
    
    \caption{\label{fig:QAOA_ave}%
        Benchmarking Momentum-QNG together with QNG, Momentum and Adam on the Minimum Vertex Cover problem. The vertical axis shows the mean (symbols) and the standard deviation (shaded regions) of the quality ratio (a) ($N=4$), (c) ($N=8$) and of the number of steps to convergence (b) ($N=4$), (d) ($N=8$) in a series of 200 trials, while the horizontal axis shows the learning rate $\eta$ of four different optimizers under consideration. 
        }
\end{figure}

From Fig.~\ref{fig:QAOA_ave}(a) for $N=4$ one can see that Adam, Momentum and Momentum-QNG achieve almost equal maximal values of the quality ratio within their convergence domains and significantly outperform the momentumless QNG. From Fig.~\ref{fig:QAOA_ave}(c) for $N=8$ one can see that Momentum-QNG performs very similar to Adam within its convergence domain and achieves almost the same maximal quality ratio as Momentum. At the same time, Adam achieves the highest quality ratio within its convergence domain. Again, the momentumless QNG achieves the least quality ratio.

To study the convergence behaviour of the optimization algorithms under consideration, in Fig.~\ref{fig:QAOA_ave}(b) and (d) we plot the mean (symbols) and the standard deviation (shaded regions) of the number of steps to convergence, as a function of the learning rate $\eta$. For both $N=4$ and $N=8$, Momentum demonstrates the narrowest convergence domain, QNG and Momentum-QNG exhibit similar intermediate-range convergence domains, while Adam shows the widest convergence domain.
For further details of our calculations see our code \cite{Borbysh_cover}.

\section{Conclusions} \label{conclusions}

In this paper we demonstrate that application of Langevin dynamics with Quantum Natural Gradient force for optimization of variational quantum circuits gives a new optimization algorithm, which we call Momentum-QNG. 

The basic QNG algorithm uses the Quantum Geometric tensor to rescale the variational parameter space to give a more symmetric shape of the objective function. 
On the other hand, the momentum (inertial) term in the Momentum algorithm prevents the optimization process from sticking to the local minima and plateaus. Indeed, from Eq.~\eqref{eq:sigma} one can see that the mean jump length in the vicinity of a local extremum of the objective function increases with increasing momentum coefficient. This feature allows momentum-amended algorithms to explore a wider volume in the variational parameter space and to find deeper minima.

This conclusion is supported by our numerical experiments. Indeed, for both the Investment Portfolio (see Fig.~\ref{fig:Portfolio}(a), (c) and (e)) and the Minimum Vertex Cover (see  Fig.~\ref{fig:QAOA_ave}(a) and (c)) optimization problems the momentum-amended algorithms (Adam, Momentum and Momentum-QNG) outperform the momentumless QNG. It is worth noting that Adam demonstrates the best performance in both these problems. For the quantum Sherrington-Kirkpatrick model with relatively strong transverse field $g=0.1$, the basic QNG algorithm demonstrates the best optimization performance (Fig.~\ref{fig:SK}(a)), in agreement with Stokes et al. \cite{Stokes_2020}. For $g=10^{-3}$ the Momentum and Momentum-QNG algorithms show almost equal best results (Fig.~\ref{fig:SK}(c)) and for $g=10^{-5}$ Momentum-QNG outperforms the rest (Fig.~\ref{fig:SK}(e)). One should take into account that at small values of the transverse field the spin-glass features of the quantum Sherrington-Kirkpatrick model, including multiple local minima of the energy landscape, become more pronounced. It is tempting to assume that a synergetic effect of application of the quantum geometric tensor and momentum results into this enhanced performance.

\bibliography{refs}

\section*{Acknowledgements}

This study was initiated during the QHACK bootcamp organized by HAIQU and continued during QHACK2024 organized by Xanadu Quantum Technologies. We deeply appreciate the inspiring atmosphere of both these events and further seminars and discussions. We are grateful to Dr. Mykyta Bulakhov for his help with multiprocessor cluster calculations. O.B., M.B. and I.O. acknowledge funding through the EURIZON project, which is funded by the European Union under grant agreement No.871072. I.L. acknowledges support by the IMPRESS-U grant from the US National Academy of Sciences and Office of Naval Research Global via STCU project No. 7120 and the IEEE program ‘Magnetism for Ukraine 2025’, Grant No. 9918. A.S. acknowledges support by the National Research Foundation of Ukraine, project No. 0124U004372.

\section*{Author contributions statement}

\textbf{Oleksandr Borysenko:} Writing – original draft, Methodology, Conceptualization, Investigation, Funding acquisition, Project administration. \textbf{Mykhailo Bratchenko:} Writing – original draft, Investigation, Methodology, Software, Visualization. \textbf{Ilya Lukin:} Writing – original draft, Investigation, Methodology, Software, Visualization. \textbf{Mykola Luhanko:} Writing – original draft, Investigation, Software, Visualization. \textbf{Ihor Omelchenko:} Writing – original draft, Investigation, Software, Visualization. \textbf{Andrii Sotnikov:} Writing – original draft, Investigation, Funding acquisition, Supervision, Project administration. \textbf{Alessandro Lomi:} Writing – original draft, Investigation, Supervision.

\section*{Declaration of competing interest}

The authors declare that they have no known competing financial interests or personal relationships that could have appeared to influence the work reported in this paper.

\section*{Data availability}

Our open source code and more results are available at \url{https://github.com/borbysh/Momentum-QNG}

\end{document}